# BUILDING A HEALTH CARE DATA WAREHOUSE FOR CANCER DISEASES

Dr.Osama E.Sheta[1] and Ahmed Nour Eldeen [2]

[1,2]Department of Mathematics Faculty of Science,
Zagazig University, Zagazig, Elsharkia, Egypt.
[1]`oesheta75@gmail.com`, [2]`ahmednour_cs@yahoo.com`

*Abstract*:

*This paper presents architecture for health care data warehouse specific to cancer diseases which could be used by executive managers, doctors, physicians and other health professionals to support the healthcare process. The data today existing in multi sources with different formats makes it necessary to have some techniques for data integration. Executive managers need access to Information so that decision makers can react in real time to changing needs. Information is one of the most factors to an organization success that executive managers or physicians would need to base their decisions on, during decision-making. A health care data warehouse is therefore necessary to integrate the different data sources into a central data repository and analysis this data.*

## 1. INTRODUCTION

The healthcare industry is one of the world's largest, fastest-developing and most information-rich industries [1].The growing need for integrated healthcare has led this industry to open towards adoption of extensive healthcare decision support systems. Information technology in health care is still a topical subject, has stimulated developments in physician order entry, decision support systems and shared patient records. Despite all the efforts, many health care organizations still have stand-alone information systems that do not communicate with each other. More importantly, clinical information systems (CIS) such as electronic patient records, are often designed to support hands-on care for individual patients, but are not well suited for analyses on an aggregated level, for example on groups of patients with the same disease[2].

In this paper we work in the cancer diseases, the cost of treatment for these diseases, Death rate in specific type of cancer and the impact of a particular drug on the disease.

Cancer known medically as a malignant neoplasm, is a broad group of various diseases, all involving unregulated cell growth. In cancer, cells divide and grow uncontrollably, forming malignant tumors, and invade nearby parts of the body. The cancer may also spread to more distant parts of the body through the lymphatic system or bloodstream. Not all tumors are cancerous. Benign tumors do not grow uncontrollably, do not invade neighboring tissues, and do not spread throughout the body. Determining what causes cancer is complex. Many things are known to increase the risk of cancer, including tobacco use, certain infections, radiation, lack of physical activity, poor diet and obesity, and environmental pollutants. These can directly damage genes or combine with existing genetic faults within cells to cause the disease. Approximately five to ten percent of cancers are entirely hereditary.

People with suspected cancer are investigated with medical tests. These commonly include blood tests, X-rays, CT scans and endoscopy [3].

## 2. DATA WAREHOUSE ARCHITECTURE

The concept of "data warehousing" arose in mid 1980s with the intention to support huge information analysis and management reporting [4]. Data warehouse was defined According to Bill Inmon a "subject-oriented, integrated, time variant and non-volatile collection of data in support of management's decision making process" [5].

According to Ralph Kimball "a data warehouse is a system that extracts, cleans, conforms, and delivers source data into a dimensional data store and then supports and implements querying and analysis for the purpose of decision making".[6]

Today, data warehouses are not only deployed extensively in banking and finance, consumer goods and retail distribution and demand-based manufacturing, it has also became a hot topic in noncommercial sector, mainly in medical fields, government, military services, education and research community etc.

Data warehouse architecture is a description of the components of the warehouse, with details showing how the components will fit together [7]. Figure 1 shows a typical architecture of a data warehouse system which includes three major areas that consist of tools for extracting data from multiple operational databases and external sources for cleaning, transforming and integrating this data; and loading data into the data warehouse [8].

Data are stored and managed in the warehouse and data marts which present multidimensional views of data to a variety of front end tools: query tools, report writers, analysis tools, and data mining tools [8].

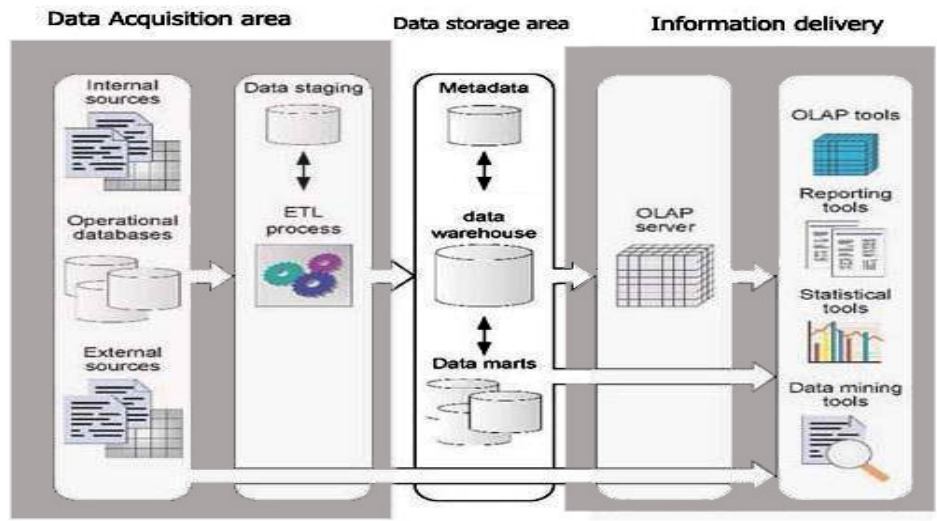

Figure 1.  Data Warehouse Architecture

There are three major areas in the data warehouse architecture as following:

- Data acquisition.
- Data storage.
- Information delivery.

Then we will explain in details each area as following: [9]

- **Data Acquisition**: This stage covers the process of extracting data from the multi sources, moving all the extracted data to the staging area, and preparing the data for loading into repository. The two major architectural components of this area are source data and data staging which is the place where all the extracted data is put together and prepared for loading into the data warehouse.

    The function and services for this area are:

    1. **Data Extraction**: Select data sources from multi sources and determine the types of data .
    2. **Data Transformation**: transform extracted data to data for data warehouse repository. Clean data, de-duplicate, and merge. De-normalize extracted data structures as required by the dimensional model of the data warehouse

- **Data storage**: This stage covers the process of loading the transformed data from the staging area into the data warehouse repository.

    The function and services for this area are the following:

    1. Load a data into data warehouse tables.
    2. Optimize the loading process.

- **Information delivery**: The information delivery component makes it easy for the users to access the information directly from the data warehouse.

    The function and services for this area are the following:

    1. Allow users to browse data warehouse content.
    2. Enable queries of aggregate tables for faster results and analysis.
    3. Provide multiple levels of data granularity.
    4. Easy to perform complex analysis using online analytical processing (OLAP).

## 3. BUILDING CANCER DATA WAREHOUSE

Many healthcare processes involve a series of patient visits or a series of outcomes. The modeling of outcomes associated with these types of healthcare processes is different from and not as well understood as the modeling of standard industry environments. The process of cancer patient in healthcare can be thought of as a value circle the center of this circle are data related to patient treatment [10, 11]. The treatment is measured or generated by all the processes and organizations around the circle show in Figure 2. This is quite different from typical processes in other industries, which usually follow a linear chain model in which a product moves through a series of steps from raw material to finished goods or from customer order to delivery [10].

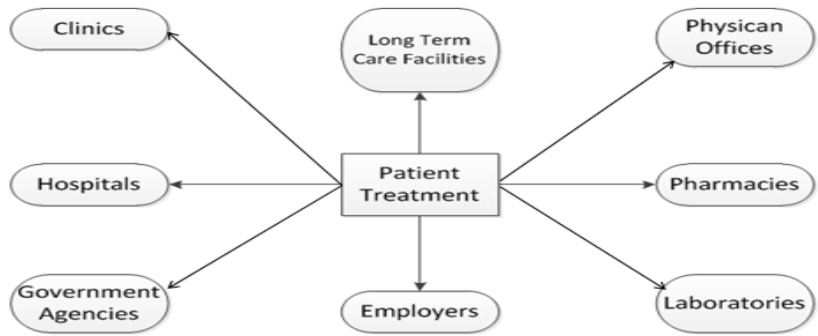

Figure 2. health care value circle

In this paper we proposes a 2 stages approach for the building cancer data warehouse

- Business Analysis.
- Architecture Design.

We will explain in details each stage as following:

## 3.1 Business Analysis

Is the discipline of identifying business needs and determining solutions to business problems. Solutions often include a systems development component, but may also consist of process improvement, organizational change or strategic planning and policy development [12].

The business analysis stage, consist of business process analysis and business requirement analysis

### 3.1.1 Business process analysis

Cancer data warehouse use case diagram as shown in Figure 3 depicts a high level overview of system functionality [13].

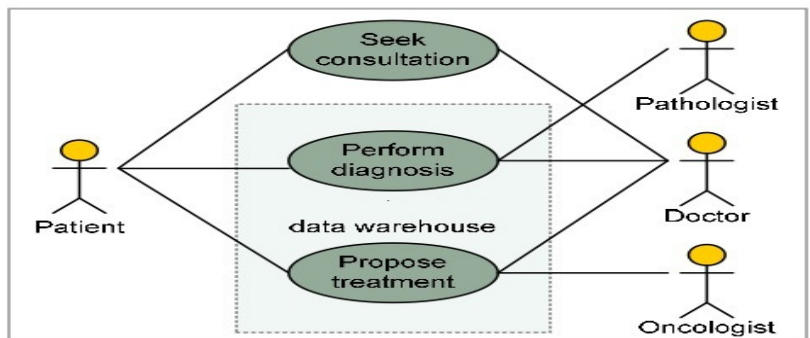
Figure 3. Cancer data warehouse use case diagram

There are four actors in the process; Patient, Doctor, Pathologist, and Oncologist. The interactions between the actors and cases are demonstrated in the following paragraphs.

Case 1: Seek consultation
Patient go to doctor for consultation when certain symptoms are noticed.

Case 2: Perform diagnosis
The doctor together with pathologist will perform a series of tests on the patient to determine the type, stage and prognosis of the cancer.

 Case 2.1: Determine the type of Cancer
First, the doctor will perform an excisional biopsy, by taking tissue sample from the affected organs for further examination. If other organs e.g. skin, brain, stomach are affected, a biopsy from these organs is also required.  Next, pathologist will inspect the physical appearance of the biopsy sample under a microscope, or identify the special molecules on the Cancer cells using markets that result to determining the type of Cancer.

Case 2.2: Determine the stage of Cancer
The doctor will proceed with a number of tests to see how advanced the cancer is and how far it has spread.

 Case 2.3: Determine the prognosis of Cancer
Next, the doctor will look into other factors to determine the prognosis of the disease.

Case 3: Propose treatment
When the above investigations are completed, the doctor and oncologist will counsel the patient regarding the best treatment options available, based on the type and the stage of the disease and some prognostic factors. There are three main types of treatment normally used to cure Cancers:

- Chemotherapy – Using drugs as infusions into the patient's veins.
- Radiotherapy – Using high energy rays over the affected areas.
- Biological therapy or antibiotic therapy – Using drugs like Rituximab to target special molecules on the cancer cells.

Besides proposing the treatment options to patient, the doctor also need to explain to the patient about the risks of taking the particular treatment and chance of recovery. Upon endorsement by the patient, the doctor will schedule the treatments for the patient.

### 3.1.2 Business Requirement analysis

We will propose some of the standard requirements for a healthcare data warehouse to support Cancer diagnosis and treatment recommendation:

- The medical diagnosing function requires understanding of patient medical history, symptoms, drug-drug interactions, knowledge of diseases in general as well as the general population.
- Minimal level of demographic details about the patient need to be captured in the patient master record include: Full Name, Gender, Date of Birth, Marital Status, Address, Occupation and etc.
- All patient master records can be recognized by National ID number.  The system must be able to detect duplication of patient master records.  The system shall automatically merge the patient records, if duplication of master records for the same patient is found.
- The system must able to display data at both summary and detail levels, and allow users to drill down to analyze a specific result.

- The system must facilitate a multi-dimensional data model presentation. Patient and medical details can be queried by multiple factors, such as location, gender, age range, blood group, race etc.
- The system should have to be updateable and adapted to constant changes that accompany the scientific development.

## 3.2 Architecture Design

Figure 4 illustrates the overall architecture of the propose healthcare data warehouse specific to Cancer. Data is imported from several sources and transformed within a staging area before it is integrated and stored in the production data warehouse for further analysis.

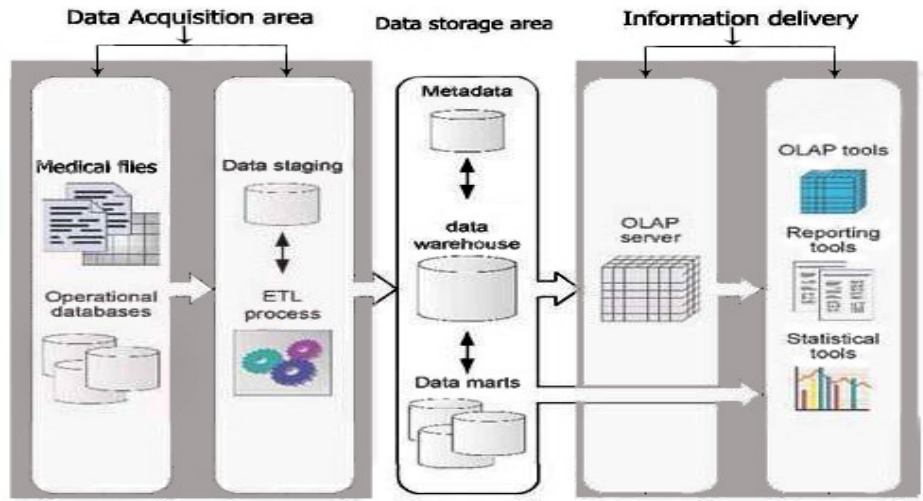

Figure 4. Cancer data warehouse Architecture

### 3.2.1 Data Acquisition

This covers the entire process of extracting data from the Access database, medical files such as (patient medical records, blood tests, urine test results, x-ray results and etc.) then moving all the extracted data to the staging area and preparing the extracted data for loading into data warehouse repository.
    In this stage there are a set of functions and services such as:

- **Data Extraction:** Select data from access DB and Medical files and determine the types of filters to be applied to Cancer data warehouse.

- **Data Transformation:** Map extracted data to data for data warehouse repository. Clean data, de-duplicate, and merge data from access DB and medical files.

### 3.2.2 Data storage:

This covers the process of loading the transformed data from the staging area into the data warehouse repository. All functions for transforming and integrating between the access data base and medical files are completed in the data staging area.

44

In this stage there are a set of functions and services such as:

- Load a health care data into data warehouse tables.
- Optimize the loading process.
- Perform incremental loads at regular prescribed intervals.
- Support loading integrated data into multiple tables.

### 3.2.3 Information delivery:

The information delivery component makes it easy for the doctors and decision making to access the information directly from the data warehouse. In the Cancer data warehouse model, after locating the sources of data, we proceeded to design the Data Warehouse; the star model was adopted, based on a fact table and multiple dimensions tables [14]. Then create hierarchy from dimension tables and specify the measure in fact table which used with this hierarchy to create useful reports. The doctors and decision makers perform analysis using the information cubes in the MDDBs.

In this stage there are a set of functions and services such as:

- Allow doctors and decision makers to browse data warehouse content.
- Enable queries of aggregate tables for faster results and analysis.
- Provide multiple levels of data granularity.
- Monitor doctors and decision making access to improve service and for future enhancements.

Finally the designed cancer data warehouse was done using the following:

- Relational database management system: Microsoft Sql Server 2008 and Microsoft Access
- ETL : Sql server integration service 2008 (SSIS).
- OLAP: Sql server analysis service 2008 (SSAS).
- Reporting: Sql server Reporting service 2008( SSRS) .
- C # language.

## 4. CONCLUSION

The healthcare industry is one of the world's largest, fastest-developing and most information-rich industries for take advantage from this information we build the Cancer data warehouse to integrate between the operational data base and medical files and therefore the analysis on data makes easy by using OLAP cubes and viewing multilevel of details from the data. Then we can analyses the cancer diseases, the cost of treatment for these diseases, Death rate in specific type of cancer and the impact of a particular drug on the disease.

During the building data warehouse we are encounter a several challenges such as the following:

- Data warehouse development requires specialized skills that are very different from a typical database development.
- Data integration plays the most critical role in a data warehouse Building.
- Data should be extracted from physical medical files, such as patient medical records, blood tests, urine test results, x-ray results, CT scan results etc., or retrieved directly from the operational medical system.

## FUTURE WORKS

We implemented this research to DSS for Health Insurance Organization – Elsharkiya Branch - Egypt which could be used by doctors, physicians and other health professionals to support the healthcare process as well as to formulate the appropriate model to improve the quality of diagnosis, treatment recommendation decision making, the impact of a particular drug on the disease this implementation contain ETL process and star schema with fact and dimension table and create data cube for multi view of data and finally create reports with charts and time series evaluation this all in one package using C# programing language.